\documentclass[twocolumn,showpacs,preprintnumbers,amsmath,amssymb,prl,superscriptaddress]{revtex4}

\usepackage{graphicx}
\usepackage{float}
\usepackage{bm}
\usepackage{color}

\begin{document}

\title{Generic model for Cooper pair splitting}
\author{H. Soller}
\affiliation{Institut f\"ur Theoretische Physik, Ruprecht-Karls-Universit\"at Heidelberg, Philosophenweg 19, D-69120 Heidelberg, Germany}

\date{\today}

\begin{abstract}
A superconductor connected to normal leads allows to generate Einstein-Podolsky-Rosen pairs by Cooper pair splitting. It has been realized with quantum dots either defined in carbon nanotubes or InAs nanowires. After establishing the presence of Cooper pair splitting in such devices new works have invesigated the effects of a finite potential difference between the quantum dots to improve and characterize the efficiency of Cooper pair splitting. In this paper we present a generic model for Cooper pair splitting and develop two minimal models specifically for the two experimental realisations and compare them to experimental data. In addition we also explore the relation of nonlocal charge transfer to positive current cross correlation of currents and discuss the temperature dependence of Cooper pair splitting.
\end{abstract}


\maketitle

\section{Introduction}

\begin{figure}[ht]
\centering
\includegraphics[width=8cm]{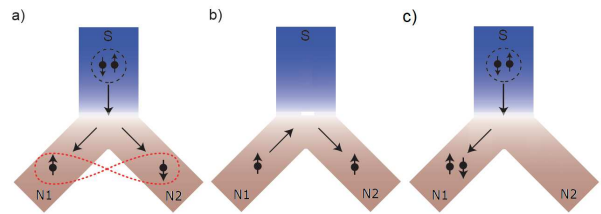}
\caption{(a) Illustration of Cooper pair splitting in a Y-junction geometry with the superconductor (S) acting as a source for entangled electrons in the drains N1 and N2. (b) Elastic cotunneling possible at finite bias between N1 and N2. (c) Local pair tunneling or Andreev reflection which transfers a pair of electrons via the same dot. Local pair tunneling refers to the break-up of a Cooper pair and the separate transfer of one electron after the other via the quantum dot whereas Andreev reflection means the simultaneous transfer of both electrons onto the same dot and then to the normal lead.}
\label{fig1}
\end{figure}

The possibility of on-chip splitting of Cooper pairs started intense theoretical \cite{PhysRevB.83.125421,PhysRevB.85.035419,PhysRevB.84.115448,0957-4484-14-1-318} and experimental effort \cite{PhysRevLett.104.026801,2009Natur.461..960H,2012arXiv1205.2455D,2012arXiv1205.1972H,2012arXiv1204.5777S}. A Cooper pair splitter consists of a superconductor contacted to two normal drains to allow for Cooper pairs to be spatially separated and transferred further to the two normal electrodes. Since the electrons in a Cooper pair form spin singlets such devices allow for on-chip generation of spin-entangled Einstein-Podolsky-Rosen pairs \cite{PhysRev.47.777,RevModPhys.81.1727} (see Fig. \ref{fig1}(a)) if the electrons can be separated coherently \cite{PhysRevB.63.165314,springerlink:10.1007/s10051-001-8675-4,PhysRevB.70.245313}. However, other competing processes like direct tunneling between the leads (elastic cotunneling) (see Fig. \ref{fig1}(b)) or transfer of both electrons to one lead (Andreev reflection or local pair tunneling) (see Fig. \ref{fig1} (c)) complicates the detection of Cooper pair splitting and reduces the device efficiency \cite{2012arXiv1204.5777S}. Recently, the first experiments at zero bias between the normal drains \cite{PhysRevLett.104.026801,2009Natur.461..960H} realized experimental control of Andreev reflection by Coulomb interaction \cite{PhysRevB.63.165314,PhysRevB.63.094515,Kicheon199936}.\\
\\
Leaving both normal drains at the same chemical potential one can study the competition between local processes like Andreev reflection or local pair tunneling with the nonlocal process of Cooper Pair Splitting (CPS). If one, however, applies a finite bias between the two normal leads the key issue is the competition between Cooper Pair Splitting (CPS) and Elastic Cotunneling (EC) in the nonlocal conductance. Theoretically the cancellation of the two contributions to the nonlocal conductance for thick tunnel barriers was shown to be removed by ferromagnetic leads \cite{0295-5075-54-2-255,PhysRevB.70.174509}, increasing transparency \cite{PhysRevB.74.214510,PhysRevB.75.172503,PhysRevB.79.104518}, or taking into account Coulomb interactions \cite{2007NatPh...3..455Y}. Also, the impact of nonequilibrium effects at high bias have been discussed \cite{PhysRevB.80.174508}. Initial experiments investigating these issues were based on nanolithographically defined metallic diffusive samples \cite{PhysRevLett.93.197003,PhysRevLett.95.027002,2009NatPh...5..393C,0295-5075-87-2-27011,2010NatPh...6..494W}, but the latest experiment on CPS at finite bias extended these questions to semiconducting nanowires \cite{PhysRevLett.107.136801}. New experiments on carbon nanotube based splitters \cite{2012arXiv1204.5777S} showed that competition between local and nonlocal processes can be directly observed. The essential difference to experiments on metallic samples is the possbility to control Andreev reflection by Coulomb interaction on the quantum dots and the possbility of local gating to identify nonlocal processes. Another control parameter is finite bias between the normal leads as was emphasized in \cite{PhysRevB.84.115448} considering a carbon nanotube based splitter, where perfect CPS is expected for an optimal bias range.\\
In this work we want to investigate Cooper pair splitters both based on carbon nanotubes and InAs. So far several models have been used in order to explain parts of the experimental data \cite{2012arXiv1205.1972H,PhysRevLett.107.136801,2012arXiv1204.5777S}. We provide a firm theoretical basis for models similar to the ones in \cite{2012arXiv1204.5777S,PhysRevLett.107.136801} and compare our results to experiment.\\
We use the generic model from \cite{PhysRevB.84.115448} for two different experimental realisations. The first one is using semiconducting nanowires where the connection between the superconductor and the normal drains is realised via two quantum dots being weakly coupled directly through the superconductor \cite{recher}. The second one is using carbon nanotubes where the connection between the two quantum dots is realised via the carbon nanotube itself which acquires a superconducting order parameter due to the proximity effect. The description of the second experimental realisation using the generic model has already been done in \cite{PhysRevB.84.115448} whereas the description of the first one will be done in this work using previous results of the $T$-matrix approach \cite{PhysRevB.63.165314}. We will compare both approaches to experiment. For the experimental approach using carbon nanotubes we will also describe how to obtain the masterequation used in \cite{PhysRevB.82.184507} in order to prove that the different nature of coupling between the quantum dots is observed experimentally. Finally we will demonstrate that the generic model is not limited to studies of conductance but can easily be generalized to higher order cumulants using full counting statistics \cite{2012arXiv1205.2455D}. Such investigation is necessary both for understanding the possible aspects of the choice of materials and for better understanding of recent experimental results \cite{PhysRevLett.107.136801,2012arXiv1205.2455D,2012arXiv1205.1972H,PhysRevLett.107.136801}.\\

The paper is organized as follows: 
in Section \ref{generic} we will review the typical experimental realization of a Cooper pair splitter and its generic model. For the generic model we will discuss the different types of coupling between the quantum dots either using a microscopic model of carbon nanotubes or the  $T$-matrix approach for semiconductor nanowires. In Section \ref{s2} we apply the $T$-matrix approach to InAs based splitters and compare its results for the finite bias situation to the experiment. Section \ref{s3} will be devoted to the analysis of carbon nanotube based splitters and the relation of the master equation approach and the generic model. 
Section \ref{s5} describes how to access the cross correlation of currents in the normal metals using full counting statistics. 
Our results are summarized in Section \ref{s6}.

\section{Generic model for double quantum dot CPS device}
\label{generic}

\begin{figure}[ht]
\centering
\includegraphics[width=9cm]{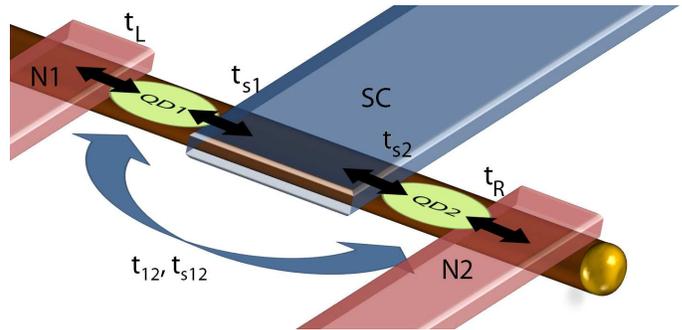}
\caption{Schematic of a generic Y-junction Cooper pair splitter. A superconductor (SC) is contacted via two quantum dots to two normal leads (L, R) at different electro-chemical potentials. The two quantum dots have tunable energy levels and can be coupled either via the superconductor ($t_{S12}$) or directly ($t_{12}$) by the host material of the QDs.}
\label{fig2}
\end{figure}

A schematic of a generic CPS device is shown in Fig. \ref{fig2}. A nanowire/carbon nanotube is used to form two quantum dots between a superconducting strip (SC) in the center and two normal metal leads (L, R). The two quantum dots can usually be tuned by two local top gates (not shown). 
\\
We use a tunneling Hamiltonian description of the system $H= H_0 + H_T$ as in Ref. \cite{PhysRevB.63.165314}, with
\begin{eqnarray}
H_0 &=& H_L + H_R + H_S + \sum_{\substack{i=1,2\\ \sigma = \uparrow, \downarrow}} \epsilon_i n_{i\sigma} \nonumber \\ 
&+& \sum_{i=1,2} U_{Di} n_{i\uparrow}n_{i\downarrow} + H_{T12},
\end{eqnarray}
and $n_{i\sigma}=d_{i\sigma}^{\dagger} d_{i\sigma}$ the occupation number operator for the dots $i=1,2$ and $H_{L,R}$ and $H_S$ Hamiltonians of the uncoupled normal (L and R) and superconducting (S) leads, respectively. The dots are characterized by spin-degenerate resonant levels at energies $\epsilon_i$ and charging energies $U_{Di}$. The inter-dot coupling $H_{T12}$ either directly or via the SC (effective tunnel rates $t_{S12}$ and $t_{12}$ in Fig. \ref{fig2}) will be introduced later as an exact double-dot Hamiltonian including the coupling with the SC electrode.  $H_T$ corresponds to the tunnel coupling between the dots and the leads,
\begin{eqnarray}
H_T &=& H_{NTL} + H_{NTR} + H_{ST1} + H_{ST2}\\
&=& \sum_{k\in L,\sigma} t_L \left(d_{1\sigma}^{\dagger}c_{k\sigma} + \mathrm{h.c.}\right) \nonumber \\
&&+ \sum_{k\in R,\sigma} t_R \left(d_{2\sigma}^{\dagger}c_{k\sigma} + \mathrm{h.c.}\right) \nonumber \\ 
&&+ \sum_{i=1,2} \sum_{k\in S, \sigma} t_{Si} \left(d_{i\sigma}^{\dagger}c_{k\sigma} + \mathrm{h.c.}\right) 
.
\end{eqnarray}
We assume that the superconducting lead is well described by the BCS theory with a superconducting gap $\Delta$ and the normal leads are characterized by a flat density of states around the Fermi level, $\rho_{L,R}(\mu)$.


The transport properties of this model can be obtained by a Green's function technique. In a combined dot-Nambu space described by spinor fields $\Psi = \left(d_{1\uparrow}, d^{\dagger}_{1\downarrow}, d_{2\uparrow}, d^{\dagger}_{2\downarrow} \right)^T$ and in the linear response regime, we can characterize the model by a retarded Green function
\begin{equation}
 \hat{G}^r \left( E \right) = \left[ E-\hat{h}_0+\hat{\Gamma}_{N}-\hat{\Sigma}^{r}(E)\right]^{-1} ,
\end{equation}
where the symbol $\hat{\cdots}$ denotes a $4\times4$ matrix in the combined Nambu-dot space. 
Hereafter we write $E$ instead of $E+i\eta$ and the limit $\eta\rightarrow0$ is taken to obtain the retarded component of the Green function. 
The Hamiltonian of the uncoupled double dot is given by
\begin{equation}
 \hat{h}_0 = \left(\begin{array}{cc}
                    \epsilon_1 & t_{12} \\ t_{12} & \epsilon_2
                   \end{array} \right) \sigma_z ,
\end{equation}
where $\sigma_{0,x,y,z}$ are the Pauli matrices in Nambu space and $t_{12}$ is the inter-dot tunneling rate. 
We describe the coupling to the normal leads by
\begin{equation}
 \hat{\Gamma}_N = \left(\begin{array}{cc}
                    i\Gamma_L & 0 \\ 0 & i\Gamma_R
                   \end{array} \right) \sigma_0 
\end{equation}
with $\Gamma_{L,R}=\pi t_{L,R}^2\rho_{L,R}(\mu)$ the tunneling rates in the wide band approximation for the spectral density of the leads. 

The self-energy $\hat{\Sigma}^{r}(E)=\hat{\Gamma}_S(E) + \hat{\Sigma}_U(E)$ includes the coupling to the superconducting electrode ($\hat{\Gamma}_{S}$) and the effect of Coulomb interactions ($\hat{\Sigma}_U$). 
The coupling to the superconducting electrode is given by
\begin{equation}
 \hat{\Gamma}_S = \left(\begin{array}{cc}
                    \Gamma_{S1} & t_{S12} \\ t_{S12} & \Gamma_{S2}
                   \end{array} \right) \left[g(E)\sigma_0 - f(E)\sigma_x\right] , \label{eq7}
\end{equation}
with $\Gamma_{S1,S2}$ the tunneling rates between the superconductor and each dot, $t_{S12}$ the inter-dot tunneling rate through the superconductor and $g(E)=-f(E)/\Delta=-E/\sqrt{\Delta^2-E^2}$ the dimensionless BCS retarded Green's functions of the uncoupled SC lead. Note, that due to the presence of $t_{S12}$ the model can in general not be mapped to a simple double-dot Hamiltonian as in \cite{PhysRevB.85.035419} which does not take into account the finite separation of tunneling points to the left and right lead. We will show below that a correct description of $t_{S12}$ is essential for reliable predictions of experimental data.

From $\hat{G}^r(E)$ the linear transport coefficients can be computed. The contribution to the zero temperature linear conductance due to local Andreev reflection processes at each dot is given by $R_{AL(R)}(E)=4|\hat{G}^r_{e1,h1(e2,h2)}(E)|^2\Gamma_{L(R)}^2$. The probability of Cooper pair splitting is $T_{CPS}(E)=4|\hat{G}^r_{e1,h2}(E)|^2\Gamma_L\Gamma_R$ and that of electron cotunneling is $T_{EC}(E)=4|\hat{G}^r_{e1,e2}(E)|^2\Gamma_L\Gamma_R$. 
Thus, the linear conductance at zero temperature at one of the normal electrodes, e.g., 1, is given by $G_1(E)=G_0[R_{AL}(E)+T_{CPS}(E)-T_{EC}(E)]$, with $G_0=2e^2/h$.

\subsection{Non-interacting case}
For the sake of clarity we will focus on the non-interacting ($U=0$) Cooper pair splitter in this work and concentrate on the consequences of the two experimental realisations since typical experiments use quantum dots on resonance where no interaction-specific effects are observed. Still, our formlism allows to include interactions as we will show below.\\
For both experimental realisations the inter-dot tunneling rate has two contributions, one given by the direct coupling between the dots, $t_{12}$, and another due to the coupling between the dots through the superconductor, $t_{S12}$. The former is usually taken as a constant parameter, while the latter is frequently ignored in the literature.

In the regime where $\Gamma_{L,R}\gg t_{12},t_{S12}$, and in the absence of electronic interactions, we find the self-energies $\Sigma_{e1,h2} \propto t_{S12}$ and $\Sigma_{e1,h2} \propto t_{12}$. Up to the energy-dependence for CPS the effective inter-dot couplings are constant. The transmission coefficients for CPS and EC are thus
\begin{eqnarray}
T_{EC}(E) &\stackrel{t_{12} \ll \Gamma_{L,R}}{\approx}& \frac{4 \Gamma_L \Gamma_R t_{12}^2}{((E-\epsilon_1)^2 + \Gamma_L^2) ((E- \epsilon_2)^2 + \Gamma_R^2)}, \nonumber\\
\label{genericec}\\
T_{CPS}(E) &\stackrel{t_{S12} \ll \Gamma_{L,R}}{\approx}& \frac{4 \Gamma_L \Gamma_R t_{S12}^2}{((E-\epsilon_1)^2 + \Gamma_L^2) ((E- \epsilon_2)^2 + \Gamma_R^2)}. \nonumber\\
\label{genericcps}
\end{eqnarray}
In this regime, the local Andreev processes are proportional to $\Gamma_{S1,S2}$ and will be dominant.

\subsection{Electronic interactions}

The quantity $\hat{\Sigma}_U$ takes into account the effect of Coulomb interactions within the dots. To the lowest order in $U$ this is given by a Hartree-Fock Bogoliubov approximation \cite{PhysRevB.63.094515} as $\hat{\Sigma}_U=U_{Di}\langle n_i \rangle \sigma_z + \Delta_i \sigma_x$, where $\Delta_i=U_{Di}\langle d_{i\uparrow}^{\dagger}d_{i\downarrow}^{\dagger} \rangle$ is the proximity effect induced order parameter in each dot. As long as Kondo correlations can be neglected, i.e., when the conditions are such that $T>T_K$, with $T_K$ the Kondo temperature, this term has the effect of reducing the amplitude of Andreev reflections on each dot by renormalizing the couplings to the superconductor. An extension of this scheme to second order in $U$ and beyond is straightforward \cite{PhysRevB.68.035105}.\\
However, since typical CPS experiments focus on QDs on resonance so that no Kondo correlations are expected, we can neglect the effect of Coulomb interaction keeping in mind that the bare position and width of the resonance in question is additionally affected by the interactions leading to a renormalised dot position and resonance width \cite{PhysRevB.83.125421}.

\subsection{Approaches to inter-dot coupling}

Next, we discuss the two approaches to the effective inter-dot coupling $t_{S12}$. The additional coupling $t_{12}$ is not mediated by the superconductor and therefore can safely be assumed to be constant as was also found using a microscopic description of a carbon nanotube in \cite{PhysRevB.84.115448} and will also turn out to be correct for the approach using semiconducting nanowires.\\
Since $t_{S12}$ does not refer to a single electron process it cannot be automatically assumed constant but one has to take into account the experimental setup. A microscopic description of $t_{S12}$ is introduced in \cite{PhysRevB.84.115448} using Green's function techniques applied to the carbon nanotube setup. In this setup the two dots are formed in the carbon nanotube and coupled via a superconducting part of the nanotube itself which acquires the superconducting gap using the proximity effect. The result for the inter-dot coupling with the superconductor is $\Gamma_{S12} = t_{S12}(E) [g(E) \sigma_0 - f(E) \sigma_x]$, which is equivalent to Eq. (\ref{eq7}). In this case $t_{S12}$ has a dependence on the energy and on the geometry of the system. However, for typical experiments we are only interested in the behavior around a single resonance so that we can safely assume $t_{S12}$ to be constant.\\
The second possibility is a setup using a semiconducting nanowire. In this case the two dots are formed in the nanowire and coupled directly via the superconductor on top of the nanowire. In this case we use a different theoretical model, the $T$-matrix approach. In this approach  conductance at small bias $U_i$ between the different leads $i = N1, \; SC,\;N2$ can be calculated via \cite{FerGoo1997}
\begin{eqnarray}
G(U_1=0) &=& \left. \frac{4 e^2}{h} \frac{\partial}{\partial \mu_1} |\langle f |T(\epsilon_i) |i\rangle|^2,\right|_{U_1=0} \; \mbox{where}\\ 
T(\epsilon_i) &=& H_T \frac{1}{\epsilon_i + i \eta - H} (\epsilon_i - H_0), \label{tmatrix}
\end{eqnarray}
is the on-shell transmission or $T$-matrix, with $\eta$ being a small positive real numer that we take to zero at the end of the calculation and we assumed to measure the conductance at bias $U_1=0$ between the SC and the first normal lead. If we would measure the conductance at a different voltage we would simply have to replace $\epsilon_i \rightarrow \epsilon_i + \mu_1$. The $T$ matrix can be written as a power series in the tunnel Hamiltonian $H_T$
\begin{eqnarray}
T(\epsilon_i) = H_T + H_T \sum_{n=1}^\infty \left(\frac{1}{\epsilon_i + i \eta - H_0} H_T\right)^n.
\end{eqnarray}
We approach the nonlocal conductance properties of the double dot Cooper pair splitter in two steps. First we consider the limit of small coupling to the SC and small bias. Later we will show that these restrictions can be relaxed considerably to also allow predictions for experimental data. In \cite{PhysRevB.63.165314} this calculation has been performed for the initial and final state of CPS in the regime 
\begin{eqnarray}
\Delta, \; U_{1,2} > e U_{N1,\;N2,\;SC}, \Gamma_l, \; k_B T > \Gamma_{Sl}. \label{reqtmatrix}
\end{eqnarray}
The result for the conductance has the form 
\begin{eqnarray}
G'_{CPS} &=& \frac{4e^2}{h} \frac{4 \Gamma_{S1} \Gamma_{S2} \Gamma_L \Gamma_R}{[(\epsilon_1 + \epsilon_2)^2 + (\Gamma_L + \Gamma_R)^2]^2} \nonumber\\
&& \times \left[\frac{\sin (k_F \delta r)}{k_F \delta r}\right]^2 \exp \left(- \frac{2 \delta r}{\pi \xi}\right), \label{gcps}
\end{eqnarray}
where $\delta r = r_1 - r_2$ and $\Gamma_{Sl} = \pi \rho_{0S} |t_{Sl}|^2$ are the single electron tunneling rates between the dots and the SC. The DOS $\rho_{0S}$ is energy independent since we assume the bias to be small. $k_F$ is the Fermi velocity in the SC and $\xi$ is the SC coherence length. Mind that for our result we use energy conservation as in \cite{PhysRevB.63.165314} but our result does not involve a second integral over energies since we calculate the conductance and not the current.\\
For EC the problem is spin symmetric and we discuss the spin-$\uparrow$ case. The final state for EC from N1 to N2 has the form $|f\rangle = |a_{1p\uparrow}a_{2q\uparrow}^+ |i\rangle$. Exchanging $1$ and $2$ in this expression gives the opposite direction for EC which is related to our result by symmetry. We know from above that the relevant tunnel coupling between the dots via the SC is small and tunneling can therefore be treated to second order in perturbation theory. Tunneling between the leads and the dot can, however, be resonant and we therefore need to treat tunneling to all orders in $H_{NTl}$. Proceeding along the lines of \cite{PhysRevB.63.165314} we can write the transition amplitude between intial and final state as
\begin{eqnarray}
\langle f |T(\epsilon_i) | i \rangle = \langle a_{1p\uparrow} T_L' d_{1\uparrow}^+\rangle \langle d_{1\uparrow} d_{2\uparrow}^+ T'' \rangle \langle d_{2\uparrow} T'_R a_{2q\uparrow}^+\rangle, \label{tex}
\end{eqnarray}
where the partial $T$-matrices $T'_l$ and $T''$ are given by
\begin{eqnarray}
T'' &=& \frac{1}{i \eta - H_0} H_{ST1} \frac{1}{i \eta - H_0} H_{ST2} \frac{1}{i \eta - H_0}, \label{t2}\\
T'_l &=& \sum_{n=0}^\infty \left( \frac{1}{i \eta - H_0} H_{NTl}\right)^{2n}.
\end{eqnarray}
We first consider the tunnel process in the superconductor. The emerging integral over the contact area can be taken from Ref. \cite{springerlink:10.1140/epjb/e2003-00361-6}. Summing over the electron and hole contribution we obtain
\begin{eqnarray}
\langle d_{1\uparrow} d_{2\uparrow}^+ T'' \rangle = \frac{\pi \rho_{0S} t_{S1} \gamma_{S2}}{(\epsilon_1 - i \eta) (\epsilon_2 - i \eta)} \frac{\cos(k_F \delta r)}{k_F \delta r} e^{- \delta r/\pi \xi}.
\end{eqnarray}
The other two matrix elements in Eq. (\ref{tex}) can be taken from Ref. \cite{PhysRevB.63.165314}
\begin{eqnarray}
\langle a_{1p\uparrow} T_L' d_{1\uparrow}^+\rangle &=& \frac{\epsilon_1 - i \eta}{\epsilon_1 - i \Gamma_L},\\
\langle d_{2\uparrow} T'_R a_{2q\uparrow}^+\rangle &=& \frac{\epsilon_2 - i \eta}{\epsilon_2 - i \Gamma_R}.
\end{eqnarray}
Finally, we obtain the EC conductance
\begin{eqnarray}
G'_{EC} &=& \frac{4e^2}{h} \frac{4 \Gamma_{S1} \Gamma_{S2} \Gamma_L \Gamma_R}{(\epsilon_1^2 + \Gamma_L^2) (\epsilon_2^2 + \Gamma_R^2)} \nonumber\\
&& \times \left[\frac{\cos (k_F \delta r)}{k_F \delta r}\right]^2 \exp \left(- \frac{2 \delta r}{\pi \xi}\right). \label{gec}
\end{eqnarray}
We can reproduce Eqs. (\ref{gcps}) and (\ref{gec}) using Eqs. (\ref{genericcps}) and (\ref{genericec}) choosing
\begin{eqnarray}
t_{12}^2 &=& \Gamma_{S1} \Gamma_{S2} \left[\frac{\cos(k_F \delta r)}{k_F \delta r}\right]^2 \exp\left(- \frac{2\delta r}{\pi \xi}\right),\\
t_{S12}^2 &=& \{\Gamma_{S1} \Gamma_{S2} [(\omega - \epsilon_1)^2 + \Gamma_L^2][(\omega - \epsilon_2)^2 + \Gamma_R^2]\} \nonumber\\
&& /\{[(\omega - \epsilon_1 - \epsilon_2)^2 + (\Gamma_L + \Gamma_R)^2]^2\} \nonumber\\
&& \times \left[\frac{\sin (k_F \delta r)}{k_F \delta r}\right]^2 \exp \left(\frac{-2\delta r}{\pi \xi}\right).
\end{eqnarray}
This result shows that $t_{12}$ is constant as expected since $\delta r$ cannot be varied. However, $t_{S12}$ has a specific dependence on the sum of the two level positions of the dots for a semiconductor based splitter.

\section{Finite bias experiments in InAs} \label{s2}

\begin{figure}[ht]
\centering
\includegraphics[width=9cm]{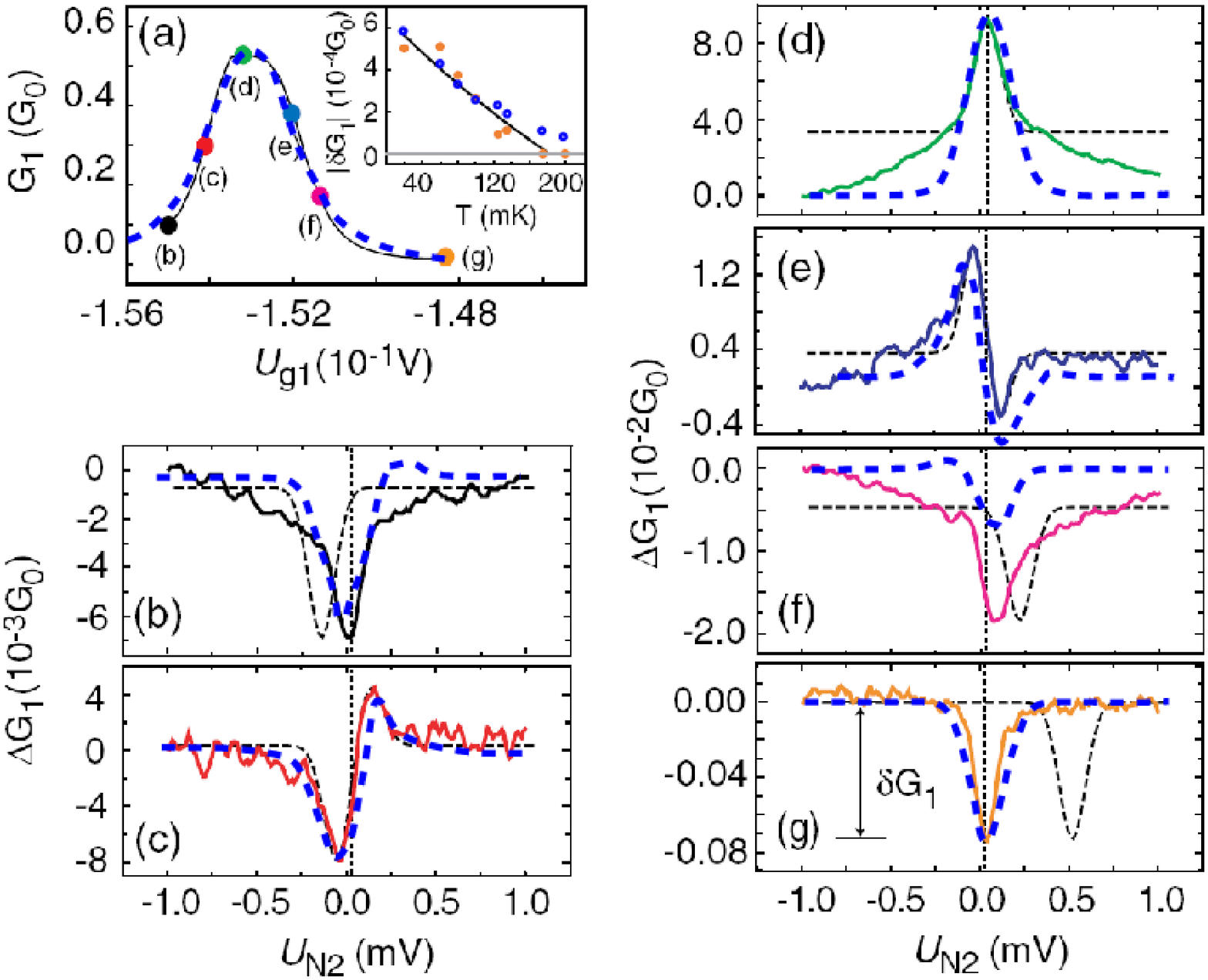}
\caption{Bias dependence of $\Delta G_1$ for a series of top-gate voltages $U_{g1}$ at $T=20$mK. The latter are indicated in (a) where $G_1$ is shown and compared to the fit plotted for $\Gamma_L = 0.12$meV, $\Gamma_{S1}=0.63$meV and $\alpha = 0.3$ using Eq. (\ref{beenakker}). The dashed curves are derived from the model discussed in the paper \cite{PhysRevLett.107.136801} and the dashed blue curves are derived from the model discussed above. Vertical lines indicate $U_{N2} = 0$, including a small offset from the IV-converter on N2. The inset in (a) shows the temperature dependence of a minimum close to the one in (g) with a black line as a guide to the eye. The blue dots are the result of the theoretical model. We use parameters $M_{CPS} = 0.195$, $M_{EC} = 0.095$, $\Gamma_{CPS} = 0.25$meV, $\Gamma_{EC1} = 0.21$meV and $\Gamma_{EC2} = 0.27$meV.}
\label{fig3}
\end{figure}

In the finite bias measurement in \cite{PhysRevLett.107.136801} first the local zero bias conductance $G_1$ between the superconductor (SC) and the normal lead (N1) is measured as a function of gate voltage around a resonance (see Fig. \ref{fig3}(a)). Then the voltage $U_{N2}$ applied on N2 relative to N1 and the SC is varied. The main contribution to the conductance comes from local transport processes between the SC and N1. To visualize the nonlocal processes one has to note that all local transport processes at N1 are independent of the voltage $U_{N2}$ on N2. Assuming that all nonlocal processes become ineffective at $U_{N2} \gg \Delta/e$ the nonlocal conductance $\Delta G_1$ is obtained from the experimental data as: $\Delta G_1(U_{g1}, U_{N2}) = G_1(U_{g1}, U_{N2}) - G_1(U_{g1}, U_{N2} = - 1mV)$.
\\
In the following we want to show how to use the generic model in order to explain the experiment as well as how to build a simpler model from the $T$-matrix approach specifically for semiconductor based setups. Especially this allows to observe the specific dependence on the sum of the dot level positions.

\subsection{Simple approach for semiconductor devices}

The first observation from the experimental data is that varying the bias on the second dot leads only to small variations in the local conductance. We desribe this feature by the dot energy level $\epsilon_2$ being pinned to the chemical potential $U_{N2}$ of the second lead. Such pinning can either be caused by a Kondo resonance or simply by a small gate capacitance compared to the capacitance of source and drain.\\
In addition we have to overcome the strict requirements in Eq. (\ref{reqtmatrix}). We can immediately drop the restriction of $\Gamma_{L,R} > \Gamma_{Sl}, \; l = 1,2$ since the relevant tunnel rates are $\Gamma'_{Sl} = \Gamma_{Sl} \frac{\sin (k_F \delta r)}{k_F \delta r} \exp\left(- \frac{\delta r}{\pi \xi}\right)$ for CPS and $\Gamma''_{S1,2} = \Gamma_{S1,2} \frac{\cos (k_F \delta r)}{k_F \delta r}\exp \left(- \frac{\delta r}{\pi \xi}\right)$ for EC, which are small even if $\Gamma_{Sl}$ is not due to the geometric suppression, especially in view of $\delta r \approx 200 \mathrm{nm}$ in a typical experiment \cite{2009Natur.461..960H}.\\
Finally, we relax the second restriction of a small bias $U_{N2} < \Delta/e$. Manifold works adress the effects of the SC DOS at finite bias \cite{PhysRevLett.80.2913,PhysRevLett.87.067006,PhysRevB.59.1637,PhysRevB.82.184507} which basically leads to a replacement of the simple factor $\Gamma_{S1}$ by an energy $\omega$-dependent effective tunnel rate $\tilde{\Gamma}_{S1} = \Gamma_{S1} \Delta / \sqrt{\omega^2 - \Delta^2}$. For the nonlocal processes analysed here due to the geometric suppression $\Gamma'_{Sl} \ll \Gamma_{L,R}$. In this limit of small transparency we can use a semiconductor model adapted from \cite{tinkham}, where the finite-bias transmission coefficient is given by the zero-bias transmission coefficient multiplied by the density of states of Cooper pairs. In the simplest approximation this is just a constant below and zero above the SC gap \cite{1998PhyE....2..887K}, which amounts to multiplying the zero-bias transmission coefficient by a step-function at the SC gap
\begin{eqnarray}
\rho(x) &=& \theta(\Delta-x) \theta(\Delta+x), \label{cutoff}
\end{eqnarray}
This approximation is compared to an exact solution in Appendix \ref{approx}. It applies since the width of the resonance is basically determined only by the tunnel rates to the normal leads $\Gamma_L$ and $\Gamma_R$ while the SC tunnel rates only contribute an overall scaling and become ineffective for voltages larger than the gap.\\
This way we can propose a model for the experimental data which is as simple as the heuristic model used in \cite{PhysRevLett.107.136801}.\\
Reconcilation of the geometrical suppression in Eq. (\ref{gcps}) is difficult \cite{2009Natur.461..960H}, especially in view of many possible effects that could alter the exact form of the suppression \cite{PhysRevB.65.165327,springerlink:10.1140/epjb/e2003-00361-6,2007NatPh...3..455Y}. We do not want to discuss this issue in detail and absorb it in a fitting parameter ($M_{CPS}$) to arrive at the final form for the conductance for CPS using Eqs. (\ref{gcps}) and (\ref{cutoff}) for the experiment in question
\begin{eqnarray}
G_{CPS} &=& \frac{M_{CPS} \Gamma_{CPS}^4}{\{[\alpha(eU_{g1} + \Delta_g) + e U_{N2}]^2 + \Gamma_{CPS}^2\}^2} \nonumber\\
&& \times \rho(eU_{N2}), \label{gcpsfull}
\end{eqnarray}
including the lever arm of the top gate $\alpha$. For EC we also introduce a matrix element $M_{EC}$ and using Eqs. (\ref{gec}) and (\ref{cutoff}) arrive at
\begin{eqnarray}
&& G_{EC} = \frac{4e^2}{h} [M_{EC} \Gamma_{EC1}^2 \Gamma_{EC2}^2] / \{[\alpha^2(eU_{g1} - \Delta_g)^2 \nonumber\\
&& + \Gamma_{EC1}^2] [(e U_{N2})^2 + \Gamma_{EC2}^2]\} \rho(eU_{N2}). \label{gecfull}
\end{eqnarray}
In both cases we have included the pinning of the dot resonance of the second dot to $eU_{N2}$ and the dependence on the gate voltage $U_{g1}$. In order to compare with the experimental data we rescale $U_{g1}$ with the lever arm of the top gate $\alpha \approx 0.3$.\\
At finite temperature we additionally have to integrate using the appropriate Fermi distributions
\begin{eqnarray}
G_{CPS}(U_{N2}, T) &=& \int \frac{d\omega}{2\pi} T_{CPS} [n_2(\omega) - n_F(\omega)], \label{tcps}\\
G_{EC}(U_{N2}, T) &=& \int \frac{d\omega}{2\pi} T_{EC} [n_2(\omega) - n_F(\omega)], \label{tec}
\end{eqnarray}
where $T_{EC}$ and $T_{CPS}$ are given by Eqs. (\ref{gcpsfull}) and (\ref{gecfull}), using the substitution $eU_{N2} \rightarrow \omega$ and $n_2(\omega) = n_F(\omega- eU_{N2})$.\\
The nonlocal conductance is now given by
\begin{eqnarray}
\Delta G_1 = G_{CPS}(U_{N2},T) - G_{EC}(U_{N2},T). \label{deltag1}
\end{eqnarray}
For the local conductance we use the zero bias conductance between a normal conductor and a SC coupled via a resonant level located at $\Delta_d$ without interactions has been calculated \cite{PhysRevB.46.12841}
\begin{eqnarray}
G_1 = \frac{4e^2}{h} \left(\frac{2 \Gamma_L \Gamma_{S1}}{\Delta_d^2 + \Gamma_L^2 + \Gamma_{S1}^2}\right)^2. \label{beenakker}
\end{eqnarray}
In the interacting case the same formula can be recovered \cite{PhysRevB.63.094515}, however, with renormalised parameters $\Delta_d$ and $\Gamma_{S1}$ involving the respective $11$- and $12$-component of the Nambu self-energy. Here we take the width and position of the resonances from fitting to the experimental data. As we only want to describe the quantum dot for voltages close to resonance we do not expect specific interaction effects like Kondo resonances \cite{Aleiner2002309} and a non-interacting description is justified keeping in mind that the width and position of the resonances is not only related to the bare tunneling rates but also to the interaction.\\
We fit the parameters to the experimental data and obtain $M_{CPS} = 0.195$, $M_{EC} = 0.095$, $\Gamma_{CPS} = 0.25$meV, $\Gamma_{EC1} = 0.21$meV and $\Gamma_{EC2} = 0.27$meV. We observe good agreement between the theoretical prediction and the experimental data with the exception of curve (f). In contrast to the model in \cite{PhysRevLett.107.136801} we correctly capture the position of all peaks and dips. We also obtain the correct peak heights without using additional offset conductances. As in the experimental work we find that $M_{CPS} \gg M_{EC}$ so that on resonance CPS is dominant. However, CPS depends on the sum of both level positions, whereas EC only depends on the product of the two Lorentzians associated to the two quantum dots. Therefore EC becomes dominant off resonance as the CPS peak is reduced outside the SC gap. These characteristics cannot be captured by a model either without $t_{S12}$ or a constant $t_{S12}$.\\
For the fitting parameters we find $\Gamma_{EC2}$ should be equal to $\Gamma_R$ being the tunnel rate between the second quantum dot and the second drain. $\Gamma_{EC2} = 0.27\mbox{meV}$ is also a reasonable value. From our model above we also concluded that $\Gamma_{CPS} =  \Gamma_L + \Gamma_R$, which is at least partially fulfilled: $0.25\mbox{meV} \approx 0.12\mbox{meV} + 0.27\mbox{meV} = 0.39\mbox{meV}$. Also, $\Gamma_{EC1} = 0.21\mbox{meV} \approx \Gamma_L = 0.12 \mbox{meV}$ follows approximately the model derived above. A broadening of the resonances, as observed here, can be attributed to resonant tunneling between the superconductor and the quantum dots which was neglected in the nonlocal processes.\\
Compared to the model in \cite{PhysRevLett.107.136801} two effects are accounted for in our model: first, we take into account the effect of interactions and the geometric suppression which both lead to different tunneling rates to the SC for the local and nonlocal processes. Second, we account for the energy-dependent SC gap leading to a suppression of the nonlocal processes for voltages larger than $\Delta$.\\
Next, we consider the temperature dependence of the conductance peaks. In addition to the Fermi distribution in Eqs. (\ref{tcps}) and (\ref{tec}) temperature also affects the SC gap which is given by the Thouless equation \cite{PhysRev.117.1256}
\begin{eqnarray}
\frac{\Delta(T)}{\Delta_0} = \tanh\left(\frac{\Delta(T) T_c}{\Delta_0 \cdot T}\right),
\end{eqnarray}
where $\Delta_0$ refers to the bare (zero-temperature) gap and $\Delta(T)$ is the gap at finite temperature. $T_c$ is the critical temperature known from BCS theory, $k_B T_C = \Delta_0 / 1.76$.\\
In the experiment we obtain $T_c \approx 850$mK. In the inset of Fig. \ref{fig3}(a) the temperature dependence of the EC dip close to the one shown in Fig. \ref{fig3}(g) is shown. Out model allows to study the temperature dependence of $\Delta G_1$ from Eq. (\ref{deltag1}) using Eqs. (\ref{tcps}) and (\ref{tec}).\\
We find a rapid decrease and broadening of the EC dip for increasing temperatures. The peak positions for the dips at different temperatures agree well with the experimental data in the inset of Fig. \ref{fig3}(a). Note that the model reproduces the linear decrease of the signal due to the broadening of the Fermi distribution. We note that the EC dip does not disappear completely in our theoretical model but it becomes very broad and shallow so that it almost disappears around the temperature $T \approx 175$ mK as in the experiment. This temperature is much smaller than $T_c$ and indeed we find that the SC gap only changes by 2\% for the temperatures considered here. The same characteristics have been observed in a different experiment \cite{2009Natur.461..960H} where the authors concluded that the nonlocal signal is not controlled by the bulk $\Delta$ alone. The model developped here gives a quantitative explanation of the temperature dependence encountered in the experiments without resorting to an induced SC gap in the host material: the nonlocal signal is mainly controlled by the temperature dependence of the distribution functions and not by the temperature dependence of the SC gap.

\subsection{Application of the generic model} \label{genmod}

Since using the $T$-matrix approach has been very successful in the above section we can find an expression for $t_{S12}$ in order to find the same expression from in the generic model. Including the lever arm of the top gate and the above mentioned pinning of the resonance in the second dot to the Fermi level we arrive at
\begin{eqnarray*}
t_{S12}(E)^2 = \frac{[\alpha^2(\omega-\epsilon_1)^2 + \Gamma_{S1}^2 + \Gamma_L^2][(\omega-\epsilon_2)^2 + \Gamma_R^2]}{[(\omega - \epsilon_1 - \epsilon_2)^2 + \Gamma_{CPS}^2]^2}t_{S12}^2,
\end{eqnarray*}
where we introduce the pinning of the second resonance to the chemical potential of the second lead by setting $\epsilon_2 = eU_{N2}$. The nonlocal conductance can be calculated as before via $\Delta G_1 (\epsilon_1, U_{N2}) = T_{CPS}(\omega = 0, \epsilon_1 = eU_{g1} - \Delta_g, \epsilon_2 = eU_{N2}) - T_{EC} (\omega = 0, \epsilon_1, \epsilon_2 = eU_{N2})$.

\subsection{Comment on the Kondo situation}

So far we have mainly described the situation of a quantum dot on resonance. We want to shortly comment on the situation when the quantum is in a typical Kondo situation as discussed also in \cite{PhysRevLett.107.136801}.\\
In this case the splitting efficiency is strongly reduced (no nonlocal conductance is observed) and we observe the typical double peak structure in the conductance with peaks at $V= \pm \Delta/e$ that has previously been observed in \cite{0957-4484-15-7-056}. The emergence of this phenomenon has been attributed to the Kondo effect: at energies below the Kondo temperature the dot spin hybridizes with the lead spin density and forms a singlet state. Therefore a perfectly transmitting channel between the quantum dot and the normal conducting lead opens up. On the other hand, transport between the superconductor and the quantum dot does not have to be perfect. Either the Kondo resonance width $\Gamma_K > \Delta$ so that the quasiparticles in the superconductor can couple to the Kondo resonance and we observe (almost) perfect conductance or $\Gamma_K < \Delta$ and the superconductor is only weakly coupled to the quantum dot. The strong reduction of the splitting efficiency and the overall conductance allows to conclude that we are in the latter case.\\
The cross correlation in this case has been investigated in \cite{Soller2011425}. There we found that the situation $\Gamma_K < \Delta$ can be described by a resonant level model with strongly asymmetric couplings \cite{Albrecht201315}. Due to the weak coupling of the superconductor to the quantum dot the splitting efficiency is strongly reduced. Therefore in principle the model developed here for the resonant case should also be applicable in a typical Kondo situation, however, with much smaller matrix elements $M_{CPS}$ and $M_{EC}$.\\
However, such mapping has to be handled with care since the effects of interaction in the Kondo situation may require more careful approaches as far as cross correlations are concerned, see e.g. \cite{PhysRevLett.98.056603}.

\section{Finite bias experiments using carbon nanotubes} \label{s3}

\begin{figure}[ht]
\centering
\includegraphics[width=9cm]{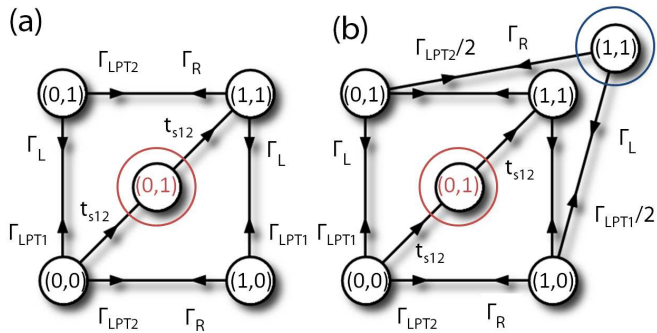}
\caption{Sketch of the different states and rates involved in solving the master equation of the carbon nanotube based Cooper pair splitter. (a) shows the scheme where only charge eigenstates are considered, whereas (b) shows the scheme when we include the difference between a triplet and a singlet state on the two quantum dots.}
\label{fig8}
\end{figure}

\begin{figure}[ht]
\centering
\includegraphics[width=8cm]{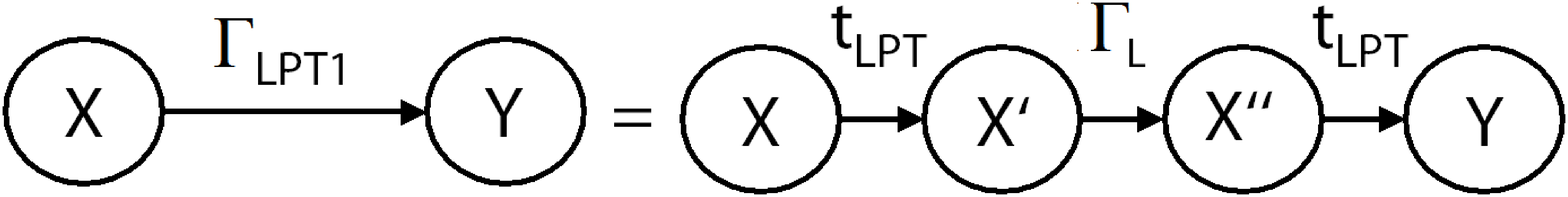}
\caption{Sktech of how to write LPT using proper rates. The example is for $\Gamma_{LPT1}$ and the treatment is analogous for $\Gamma_{LPT2}$.}
\label{fig9}
\end{figure}

In this Section we turn again to carbon nanotube based splitters as used in a recent experiment \cite{2012arXiv1204.5777S}. A microscopic model of a carbon nanotube based splitter has been presented in \cite{0953-8984-22-27-275304} where it was found that $t_{S12}$ and $t_{12}$ are constant.\\
In the experiments \cite{2012arXiv1204.5777S} it was suggested to use a master equation approach similar to \cite{PhysRevB.82.184507,PhysRevB.79.054505} to qualitatively explain the data. Here we show how to obtain the necessary rates from the generic model described in Section \ref{generic}. The approach taken in \cite{PhysRevB.82.184507,PhysRevB.79.054505} cannot be generalized in a straightforward way since the superconductor was assumed to have an infinite gap which automatically forbids processes such as local pair tunneling (LPT). We link the master equation and the generic model to determine the energy-dependence of $t_{S12}$ and $t_{12}$ that has been obtained in \cite{2012arXiv1204.5777S} in order to verify that the inter-dot coupling via the superconductor has a different energy dependence for carbon nanotube based splitters compared to semiconductor based ones.\\
In the experiment zero bias conductances between N1 and SC ($G_1$) and N2 and SC ($G_2$) have been recorded as a function of the side gate voltages on the first and second quantum dot, $U_{sg1}$ and $U_{sg2}$. Two features are observed: a typically broad local conductance peak due to LPT and a narrower additional peak due to CPS. LPT refers to the break-up of a Cooper pair due to the finite SC gap, which leads to a subsequent transfer of two electrons via one of the quantum dots. The additional CPS peaks vanish when superconductivity is suppressed using an additional magnetic field.\\
To access the nonlocal conductances $\Delta G_1$ and $\Delta G_2$ the total conductance (e.g. $G_1$) is recorded as a function of one gate voltage (e.g. $U_{sg2}$). The subtracted background is determined by manually interpolating the signal if only the local condutance would contribute. The excess conductance is then defined as $\Delta G_1$. It was suggested that such procedure corresponds to comparing $G_1$ in presence of the additional coupling $t_{S12}$ and without it, see Fig. \ref{fig2}.\\
For the masterequation we start by considering only charge eigenstates of the double dot system. In this case we have four states as shown in Fig. \ref{fig8}(a), along with the rates for the case when the leads $N1, \; N2$ have negative bias with respect to the SC. We have not included any inter-dot processes described by $t_{12}$ in the generic model, since these rates are typically small in the experiment considered \cite{2012arXiv1204.5777S}. We introduce a fifth state $(0,1)^*$ which is highlighted in red which indicates the transfer of one electron during CPS. 
\\
To lowest order in the rates for CPS, LPT and $\Gamma_L, \; \Gamma_R$ the occupation probabilities obey the master equation
\begin{eqnarray}
\sum_{\chi, \chi'} (W_{\chi\chi'} P_{\chi'} - W_{\chi'\chi} P_{\chi}) = 0,
\end{eqnarray}
where $W_{\chi\chi'}$ is Fermi's golden rule transition rate from state $\chi'$ to $\chi$, with $\chi,\chi' = (0,0), \; (0,1), \; (1,0),\; (0,1)^*, \; (1,1)$. In order to simplify notation we assume zero temperature and the leads and quantum dots gated in such a way that no backflow to the SC is possible. In this case the non-vanishing rates read
\begin{eqnarray}
W_{(0,0),(0,0)} &=& \Gamma_{LPT1}, \; W_{(0,1),(0,0)} = \Gamma_{L}, \\
W_{(0,0),(0,1)^*} &=& t_{S12}, \; W_{(0,1)^*,(1,1)} = t_{S12},\\
W_{(0,0), (1,0)} &=& \Gamma_{LPT2}, \; W_{(1,0),(0,0)} = \Gamma_{R},\\
W_{(0,1),(1,1)} &=& \Gamma_{LPT2}, \; W_{(1,1), (0,1)} = \Gamma_{R},\\
W_{(1,0),(1,1)} &=& \Gamma_{LPT1}, \; W_{(1,1),(1,0)} = \Gamma_{L},
\end{eqnarray}
The only rates we have not discussed so far are $\Gamma_{LPT1,2}$. LPT is a process which is third order in the tunnel coupling. In order to introduce proper rates we break up each arrow for LPT in Fig. \ref{fig8} as indicated in Fig. \ref{fig9}.
\\
From \cite{PhysRevB.63.165314} we immediately find
\begin{eqnarray}
t_{LPT} = t_{S12} \frac{(\Gamma_{L} + \Gamma_{R})}{\pi \Delta \left[\frac{\sin(k_F \delta r)}{k_F \delta}\right] \exp \left(- \frac{\delta r}{\pi \xi}\right)}.
\end{eqnarray}
Our model therefore gives an explicit expression of the effective rates $p_{CPS}$ and $p_{LPT}$ that were defined in \cite{2012arXiv1204.5777S}.\\
The current in lead $\eta$ can be computed from the occupation probabilities by
\begin{eqnarray}
I_\eta = \frac{e}{\hbar} \sum_{\chi \neq \chi'} W_{\chi'\chi}^\eta P_\chi,
\end{eqnarray}
where the current rates $W_{\chi'\chi}^\eta$ take into account the electrons transferred to lead $\eta$ and have to include all possible configurations in the extension of Fig. \ref{fig8} as shown in Fig. \ref{fig9}.\\
The rate $\bar{p}_{CPS}$ used for the master equation in \cite{2012arXiv1204.5777S} therefore corresponds to Eq. (\ref{genericcps}). In the experimental work the following form for $\bar{p}_{CPS}$ was used
\begin{eqnarray}
\bar{p}_{CPS}(U_{sg1}, U_{sg2}) = p_{CPS} {\cal G}_1(U_{sg1}) {\cal G}_2(U_{sg2}),
\end{eqnarray}
where ${\cal G}_i, \; i=1,2$ are Gaussian profiles in order to model the resonances of the quantum dots. Eq. (\ref{genericcps}) has the same form except for the fact that the Gaussian profiles are Lorentzians which is a better choice for a quantum dot. However, this form is clearly different from the one obtained via the $T$-matrix approach in Eq. (\ref{gcps}) and can only be obtained with a constant $t_{S12}$ as suggested from the microscopic modelling \cite{0953-8984-22-27-275304}.\\
Another remarkable feature of the experimental data is that $\Delta G_1 \neq \Delta G_2$, whereas one would naively expect $\Delta G_1 = \Delta G_2$ due to the fact that CPS is a coherent process involving both quantum dots. This discrepancy is easily reproduced when calculating the total conductance in the generic model using a sizeable $t_{S12}$.\\
The observed discrepancy can easily be explained: changing $t_{S12}$ from a finite value to zero does not only eliminate CPS but also changes the DOS on the two quantum dots. Changing the DOS also affects the local conductances so that $\Delta G_1$ and $\Delta G_2$ do not only represent CPS but also the change of the local conductances.\\
However, measuring $G_1$ and $\Delta G_1$ allows for a precise determination of $t_{S12}$ so that one may calculate the true CPS conductance from the generic model.\\
The discrepancy can easily be observed in the master equation. We choose typical parameters $t_{S12} = 0.1 = t_{LPT}, \; \Gamma_{L} = 0.12, \; \Gamma_R = 0.195$ in units of $\Delta$ and obtain for the total current $I_{1} \approx 0.043 \Delta e/\hbar$. We can also calculate the CPS current $I_{CPS} = e/\hbar \Gamma_{N1} P_{(1,1)} \approx 0.01 \Delta e/\hbar$. However, if we calculate $I_{N1}$ for $t_{S12} = 0$ we obtain $I_1(t_{S12} = 0) = 0.037 \Delta e/\hbar \neq I_{1} - I_{CPS}$. As in the generic model discussed above this is due to the different DOS of the quantum dots when $t_{S12}$ is present or not. In the master equation this change is reflected by changing probabilities $P_\chi$ which was also found in \cite{2012arXiv1204.5777S}. Again, we can extract the CPS current from fitting our model to the experimental data.\\
Two additional complications arise. First, it is not trivial to extract absolute values for the $P_\chi$, especially for the complex processes involving the SC. Second, the state $(1,1)$ is not necessarily a singlet state as it can also be populated by two consecutive LPTs with equal spins, which gives rise to a finite triplet amplitude.\\
In order to demonstrate this possibility we consider the schematic in Fig. \ref{fig8} (b), where we include a triplet state. We implement the master equation as before and calculate the probabilities, yielding
\begin{eqnarray}
\frac{P_{(1,1), \mathrm{triplet}}}{P_{(1,1), \mathrm{singlet}}} \approx 0.5,
\end{eqnarray}
so that we conclude that actually a large part is contributed by triplet splitting meaning splitting of two electrons with equal spins.\\
In conclusion we find that carbon nanotube based splitters have basic differences to semiconductor based ones. We have identified a different energy dependence of the inter-dot coupling compared to semiconductor based splitters. The difference stems from the fact that the quantum dots are coupled via the carbon nanotube and not via the SC on top. This leads to a rather strong coupling so that CPS cannot be regarded as a perturbation to the local change transport processes any longer, e.g. setting $t_{S12} = 0$ also changes the DOS of the quantum dots which causes manifold difficulties when trying to extract the exact CPS conductance \cite{2012arXiv1205.2455D}.

\section{Cross Correlation} \label{s5}

\begin{figure}[ht]
\centering
\includegraphics[width=9cm]{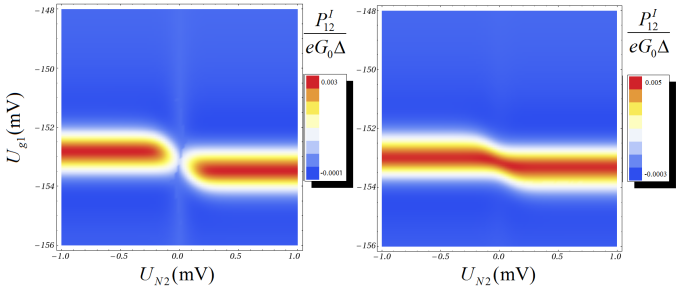}
\caption{Cross correlation of the currents in the first and second drain calculated from Eq. (\ref{cross}) using parameters $M_{CPS} = 0.195$, $M_{EC} = 0.095$, $\Gamma_{CPS} = 0.25$meV, $\Gamma_{EC1} = 0.21$meV, $\Gamma_{EC2} = 0.27$meV, $M_2 = 0.55$, $\Gamma_1 = 0.12$meV and $\Gamma_{S1}=0.63$meV as discussed before. The left plot has been calculated for $T=20$mK, the right one for $T=200$mK.}
\label{fig7}
\end{figure}

Based on the good agreement with the experiment for the InAs splitter in Section \ref{s2} we use our model to access also the cross correlation of currents. Cross correlations of this type are of interest in many systems \cite{PhysRevB.46.7061,PhysRevB.69.235305,PhysRevLett.95.146806,PhysRevB.83.201303,PhysRevLett.98.056603,PhysRevLett.93.046601}. For the system discussed here the interesting point is that for normal-conducting non-interacting fermionic systems it has been confirmed both experimentally \cite{Henny09041999} and theoretically \cite{PhysRevLett.65.2901,PhysRevB.46.12485} that the cross-correlation of currents is always negative. However, for systems with a superconducting lead it was found that due to CPS cross correlations can become positive \cite{Thierry1996137,PhysRevB.53.16390,springerlink:10.1007/s100510051010,PhysRevLett.88.197001} since the tunneling events of the two electrons of a Cooper pair are correlated. Recent experiments have confirmed this prediction \cite{2010NatPh...6..494W,2012arXiv1205.2455D}.\\
The calculation of cross correlations for realistic Cooper pair splitters involving interactions has so far been done using a Bethe-Salpeter equation \cite{PhysRevB.85.035419} since the cross correlation involves a second order correlation function. However, such approaches have so far been limited to weak interaction and have not incorporated the energy dependence of $t_{S12}$ both of which can be adressed using the generic model. The transmission coefficients calcuated from the generic model can then be implemented into a cumulant generating function to gain access to the higher order correlation function.\\
We start by introducing two transmission coefficients that describe the local processes in the SC beamsplitter. The first one is due to the local charge transfer between the SC and N1, without a bias so that $T_1 = G_1$. The second is due to local charge transfer between N2 and the SC. In the experiment the second quantum dot showed a very broad resonance \cite{PhysRevLett.107.136801}. Since we have assumed this to be due to the dot resonance being pinned to the Fermi level of the second lead, the transmission coefficient is constant. We also assume charge transfer between these two parts of the system is only due to tunneling of single electrons which is a conservative estimate as will become clear below. The amplitude of the resonance is assumed to be the same as for the first quantum dot so that we introduce
\begin{eqnarray}
T_1 = G_1/2, \; T_2 = \frac{2e^2}{h} M_2 \label{localt},
\end{eqnarray}
where $M_2$ corresponds to the height of the second resonance.\\
Having obtained all transmission coefficients we use the full counting statistics (FCS) of charge transfer, which provides direct access to all higher order cumulants via the current distribution function \cite{nazarov,nazarov-2003-35,PhysRevB.70.115305}. This procedure is by now well established and has been applied to numerous quantum impurity problems \cite{PhysRevB.82.121414,PhysRevB.82.165441,Soller2011425}. The full counting statistics for a superconducting beamsplitter have already been calculated in Ref. \cite{soller2}, however, for a slightly different geometry. Nonetheless, no other charge transfer processes but the ones already identified in \cite{soller2} are taking place so that we can use the approach presented in \cite{PhysRevLett.96.216406}: we take the structure of the FCS from Ref. \cite{soller2} and use the transmission coefficients from the model described in Section \ref{s2}. This way the cumulant generating function $\chi(\bm{\lambda})$ reads
\begin{widetext}
\begin{eqnarray}
\ln\chi(\bm{\lambda}) &=& \tau \int_{-\infty}^\infty \frac{d\omega}{\pi} \ln \left \{1+ T_1 [(e^{i 2(\lambda_1 - \lambda_s )} -1) n_F (1-n_F) + (e^{-i 2(\lambda_1 - \lambda_s)}-1) \right. \nonumber\\
&& \times n_F (1-n_F)]+  T_2 [(e^{i (\lambda_2 - \lambda_s )} -1) n_2 (1-n_F) + (e^{-i (\lambda_2 - \lambda_s)}-1) n_F (1-n_2)] \nonumber\\
&& + T_{CPS}(\omega) [(e^{i (\lambda_1 + \lambda_2 - 2 \lambda_s)}-1) n_2 (1-n_F) + (e^{-i (\lambda_1 + \lambda_2 - 2\lambda_s)}-1) \nonumber\\
&& \times n_F(1-n_2)]+ T_{EC}(\omega) [(e^{i (\lambda_2 - \lambda_1)}-1) n_2 (1-n_F) \nonumber\\
&& \left. + (e^{-i (\lambda_2 - \lambda_1)}-1) n_F(1-n_2)]\right\}, \label{cgf}
\end{eqnarray}
\end{widetext}
where the dependence on $\omega$ in the distribution function has been omitted. $\bm{\lambda} = (\lambda_1, \lambda_2, \lambda_s)$ refers to the counting fields for the first, second and superconducting lead. From the cumulant generating function we calculate the cross correlation of the currents through the first and second lead using
\begin{eqnarray}
P_{12}^I = - \frac{1}{\tau} \frac{\partial^2\ln \chi(\bm{\lambda})}{\partial \lambda_1 \partial \lambda_2}. \label{cross}
\end{eqnarray}
The result is shown in Fig. \ref{fig7} for two temperatures $T=20$ mK and $T=200$ mK. We find a positive cross correlation in a small gate voltage interval around the CPS resonance, which shows the importance of positive cross correlation as a signature of CPS. At larger temperature the positive cross correlation is enhanced due thermally excited CPS processes. For top gate voltages away from resonance we find a negative cross correlation due to EC.
\\
The general picture is therefore identical to other treatments of superconducting beamsplitters. However, we note that the overall value of the cross correlations is very small. This is due to the fact that the local processes included in the cumulant generating function in Eq. (\ref{cgf}) are involved in the denominator of Eq. (\ref{cross}) which means that they do not determine the sign but the overall value of the cross correlations. This also explains why the precise form of the transmission coefficient and the type of charge transfer for the local processes described by the transmission coefficients in Eq. (\ref{localt}) are not of prior importance. The large resonance of both transmission coefficients lead to a strong reduction of the cross correlations mediated by CPS and EC.\\
On the other hand, positive cross correlation is a direct consequence of positive nonlocal conductance due to CPS \cite{0295-5075-102-5-50009,PhysRevLett.111.136806}. Therefore as a signal of CPS nonlocal conductance is advantageous since the local processes can be neglected completely.

\section{Conclusions} \label{s6}

To conclude, we have described how to model Cooper pair splitters based on InAs and carbon nanotubes both by a generic model and by specific minimal models. We have compared our predictions to experimental data with good agreement. We identified crucial differences between semiconductor and carbon nanotube based setups. In this way we have explained several recent observations. We have also demonstrated how to access the cross correlation of currents and obtained positive cross correlation from CPS.\\
The characteristics of finite bias Cooper pair splitting are similar but not identical for carbon nanotube based and InAs nanowire based splitters. We have attributed the different behavior to the different energy dependence and magnitude of the CPS transmission coefficient. We have not observed specific effects due to interactions since the experiments have been conducted for quantum dots on resonance. From our considerations measurements of the nonlocal conductance are advantageous to measurements of the cross correlation since they reveal only the nonlocal processes.

\section*{Acknowledgements}

The author would like to thank A. Baumgartner, L. Hofstetter, K. Kang, A. Levy Yeyati, P. Burset and A. Komnik for many helpful discussions.

\appendix

\section{Approximation for effects of energy-dependent $\Gamma_{s1}$} \label{approx}

\begin{figure}[ht]
\centering
\includegraphics[width=7cm]{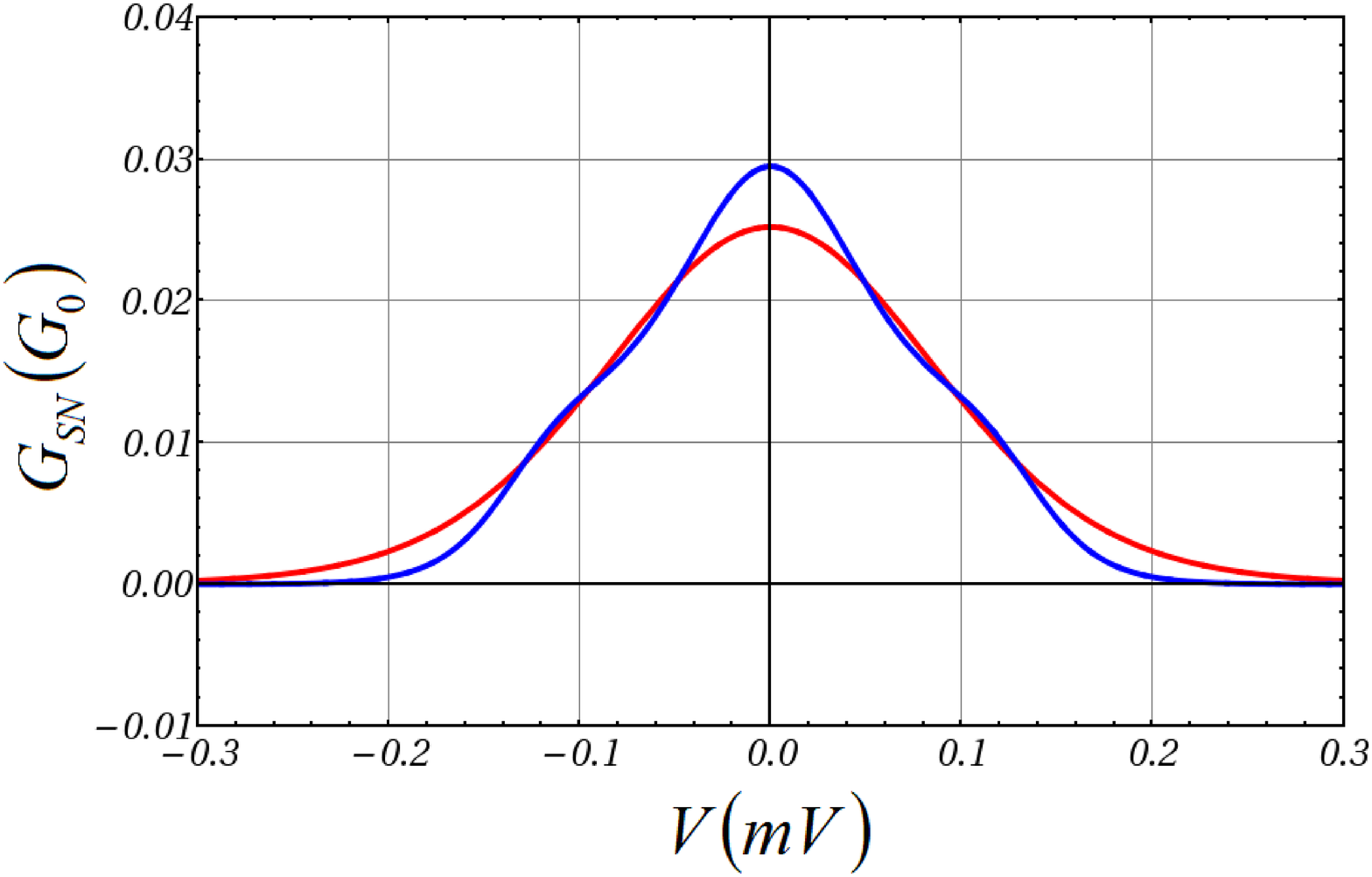}
\caption{Bias dependence of $G_{SN}$ (blue curve) taken from \cite{soller2} and $G_{SN}$ (red curve) given by Eq. (\ref{gapprox}) as a function of the applied bias $V$ between the normal metal and the SC. We use parameters $\Gamma_n = 0.2$ meV, $\Gamma_{s} = 0.01$meV, $\Delta = 0.13$meV and $T=0.02$meV in order to represent the typical situation for the nonlocal conductances. We observe that the approximate solution mostly agrees well with the exact result.}
\label{fig5}
\end{figure}

In Ref. \cite{soller2} conductance for charge transfer mediated by Andreev processes in a superconductor-quantum dot-normal metal junction in case of finite bias $V$ and finite temperature has been considered. CPS and EC are both charge transfer processes below the SC gap with rather low tunneling coupling to the SC due to the geometric suppression. Therefore a good test for the approximation of the energy dependence in Eq. (\ref{cutoff}) is to compare the approximate conductance resulting from Eq. (\ref{cutoff}) in a superconductor-quantum dot-normal metal junction at low tunnel coupling to the exact result. For convenience we consider a quantum dot with a resonance at voltage $V=0$. Eq. (\ref{beenakker}) will give the following result (for small $V$)
\begin{eqnarray}
G_{SN,\mbox{small $V$}}(V) &=& \frac{4e^2}{h} \left(\frac{2 \Gamma_n \Gamma_s}{(e^2V^2 + \Gamma_n^2 + \Gamma_s^2}\right)^2 \label{gsnbeen},
\end{eqnarray}
where $\Gamma_{n,s}$ are tunnel rates between the quantum dot and the normal metal/SC, respectively. If $eV\ll \Delta$ is not fulfilled anymore we have to include the abovementioned effects of the energy dependent SC DOS. We obtain the conductance at arbitrary $V$ from the approximation
\begin{eqnarray}
G_{SN}(V) &=& G_{SN,\mbox{small $V$}}(V) \rho(V). \label{gapprox}
\end{eqnarray}
We compare the approximation in Eq. (\ref{gapprox}) to the exact result from \cite{soller2} in Fig. \ref{fig5} for parameters typical for nonlocal conductances.
\\
We observe acceptable agreement between the exact and approximate solution.\\

\section{Finite bias splitting with constant $t_{S12}$} \label{fbiasgenmod}

We check that the generic model with a constant $t_{S12}$ does not lead to the same voltage dependence of EC and CPS and try to reproduce the behavior of the nonlocal conductance as indicated in Fig. \ref{fig3}: on resonance a peak from CPS is observed whereas off resonance this peak dissapears and turns into a dip caused by EC. Zero-bias conductance is obtained as in Section \ref{genmod}, however, with a constant $t_{S12}$.
\begin{figure}[ht]
\centering
\includegraphics[width=9cm]{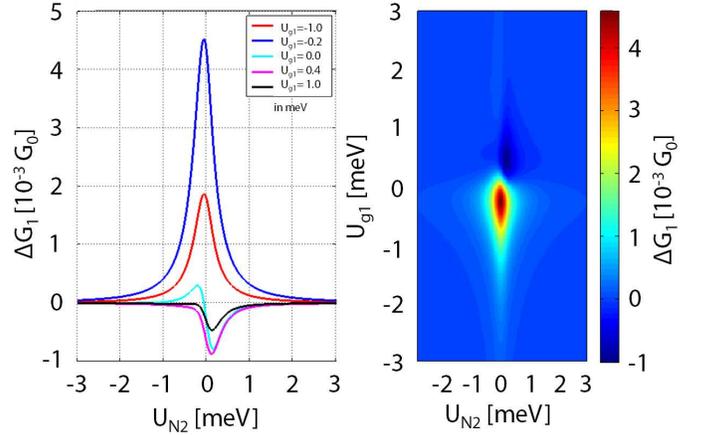}
\caption{Left: Plot of the nonlocal conductance $G_{CPS}-G_{EC}$ at zero energy and temperature as a function of the gate $V_{G2}$ for different values of $V_{G1}$ calculated using the generic model. The tunneling rates are $t_L=0.6$ meV, $t_R=0.3$ meV, $t_{S1}=0.03$ meV, $t_{S2}=0.005$ meV, $t_{S12}=0.03$ meV, $t_{12}=0.029$ meV. The superconducting gap is $\Delta=0.13$ meV. 
Right: Color map of the nonlocal conductance as a function of the gates $V_{G1}$ and $V_{G2}$ for the same parameters.}
\label{fig-GM}
\end{figure}
In Fig. \ref{fig-GM} we observe the same qualitative behavior as observed in the InAs splitter experiment, however, we have not reached quantitative agreement.

\end{document}